\documentclass{epl}
\usepackage{epsfig}

\newcommand{\be}{\begin{equation}}
\newcommand{\ee}{\end{equation}}
\shorttitle{Length scale for Fickian diffusion}
\institute{
    \inst{1}LDV UMR 5587, Universit\'e Montpellier II
and CNRS, 34095 Montpellier, France\\
    \inst{2}Rudolf Peierls Centre for Theoretical Physics,
University of Oxford, 1 Keble Road, Oxford, OX1 3NP, UK \\
    \inst{3}Department of Chemistry, University of California,
Berkeley, CA 94720-1460, USA \\
    \inst{4}School of Physics and Astronomy, Univ. of Nottingham,
Nottingham, NG7  2RD, UK
}
\shortauthor{L. Berthier \and D. Chandler \and J.P. Garrahan}

\begin{document}

\title{Length scale for the onset of Fickian diffusion in supercooled liquids}
\author{Ludovic Berthier\inst{1,2} \and David Chandler\inst{3} \and
Juan P. Garrahan\inst{4}}
\maketitle

\begin{abstract}
The interplay between self-diffusion and excitation lines in space-time was
recently studied in kinetically constrained models to explain the breakdown
of the Stokes-Einstein law in supercooled liquids. Here, we further examine
this interplay and its manifestation in incoherent scattering functions. In
particular, we establish a dynamic length scale below which Fickian
diffusion breaks down, as is observed in experiments and simulations. We
describe the temperature dependence of this length scale in liquids of
various fragilities, and provide analytical estimates for the van Hove and
self-intermediate scattering functions.
\end{abstract}

\vspace*{-.85cm} \textit{A ten-day journey starts with a single step.} ---
Laotse, Tao Te King.

\vspace*{.4cm}

\begin{figure}
\begin{center}
\epsfig{file=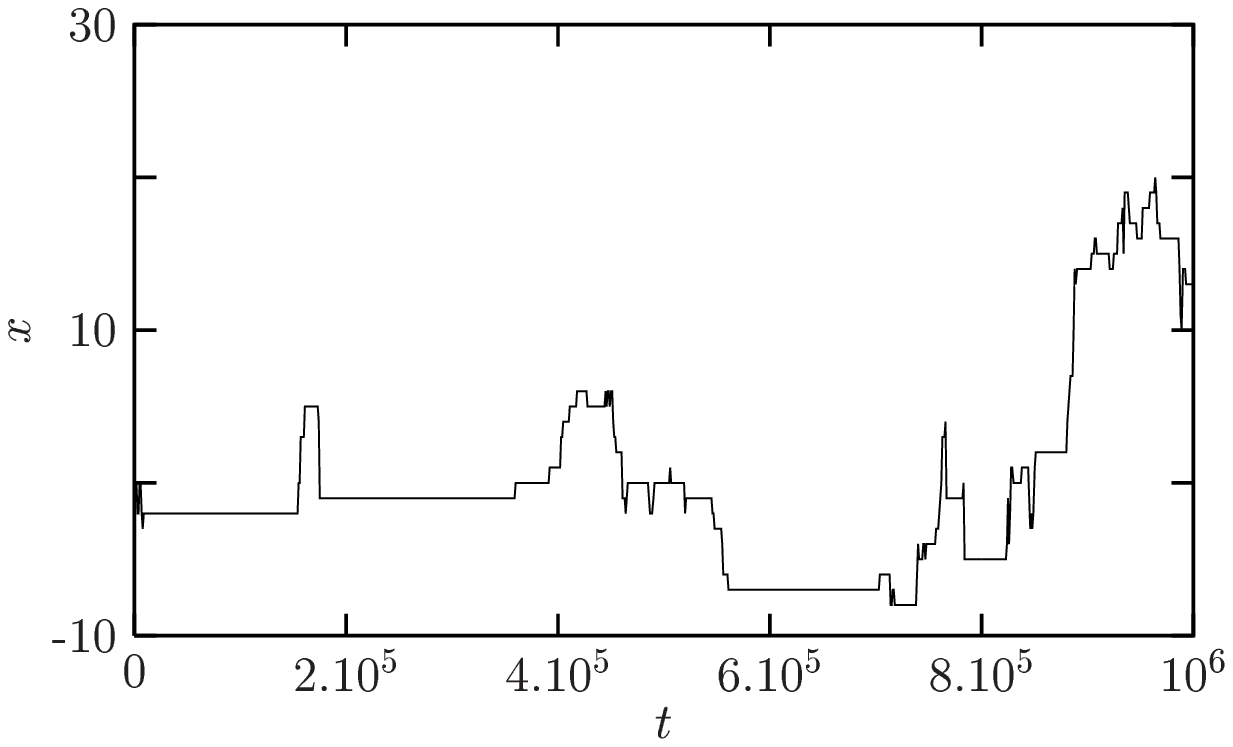,width=6.7cm} \psfig{file=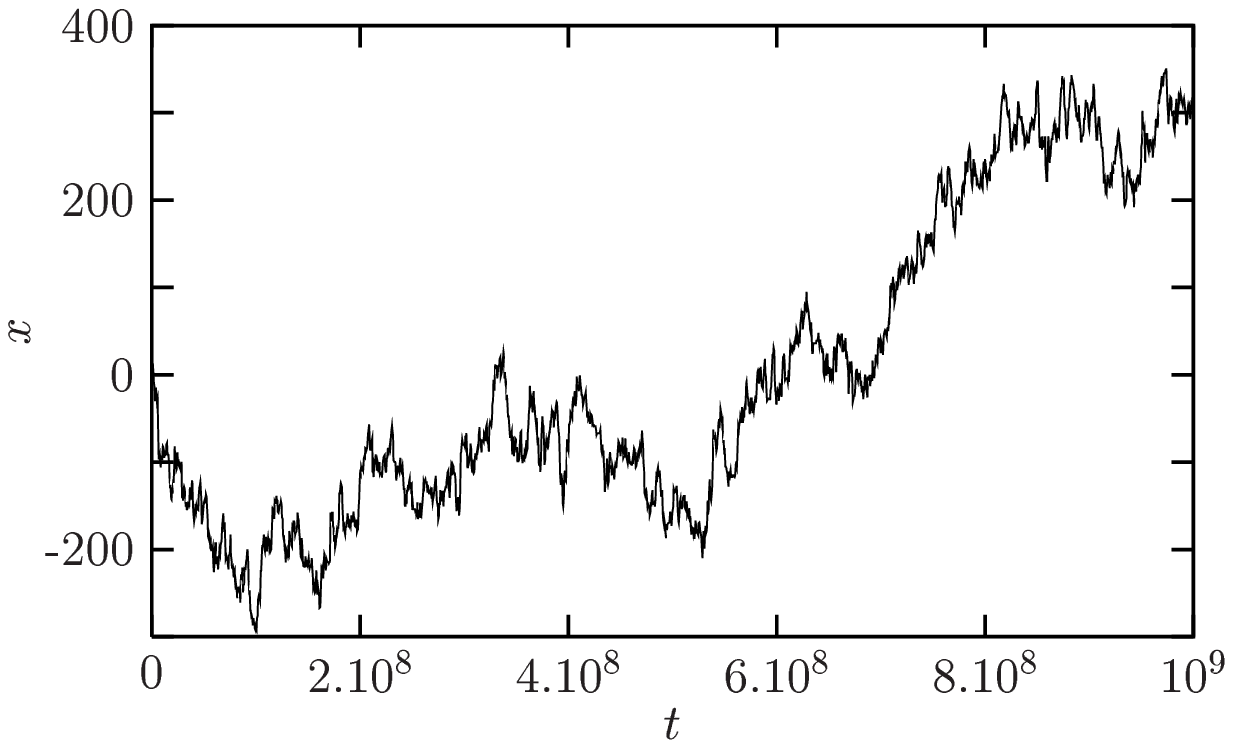,width=6.7cm}
\end{center}
\caption{A typical trajectory of a probe molecule in the 1D FA model
   at $T=0.25$ is shown twice on different time and length scales. The
   $\alpha$-relaxation time is $\tau_\alpha \sim 10^5$, the diffusion
   constant $D \sim 10^{-4}$.}
\label{bubble}
\end{figure}

In this paper, we consider the process of self-diffusion of probe molecules
in supercooled liquids. Figure~\ref{bubble} shows the trajectory of a such a
probe obtained from a model of a supercooled liquid~\cite{jung}. At
conditions shown, the structural relaxation time of the model is of the
order of $10^{5}$ microscopic time steps. The left panel of Fig.~\ref{bubble}
extends over this range of time. The right panel extends three orders of
magnitude longer in time, and one order of magnitude larger in space. Here,
the trajectory looks like a random walk of Fickian diffusion, unlike the
trajectory in the left panel. In this paper, we describe the crossover from
non-Fickian to Fickian diffusion, and identify the length scale, $\ell
^{\star }$, that characterizes the crossover.

In our perspective supercooled liquids are modeled by a 
coarse-grained mobility field evolving with simple empirical 
rules~\cite{reviewkcm}. 
An essential prediction of our modeling is that 
dynamics becomes spatially correlated~\cite{garrahan-chandler}, i.e.,
the growth of timescales is accompanied by the growth of dynamical
lengthscales, giving rise to the phenomenon of dynamic heterogeneity
observed in experiments and simulations~\cite
{DHreviews1,DHreviews2,DHreviews3,DHreviews4}. 
This finding suggests a degree of universality and thus utility for a 
coarse-grained perspective. Nevertheless, it is important to test of 
our approach by using it to revisit 
all sorts of experimental and numerical
studies of supercooled liquids~\cite{jung,pnas,pre,jcp}. 
Generic properties are only weakly
dependent upon details of the models, which become important, however,
for quantitative comparisons to experiments or
simulations~\cite{pnas}. In most of this paper we will therefore
pursue our investigations in the simplest lattice model of this
family~\cite {reviewkcm}, namely the one-dimensional
Fredrickson-Andersen model (hereafter 1D FA model)~\cite{fa}. This is
defined by the Hamiltonian $H=\sum_{i}n_{i}$, where $n_{i}=0,1$, with
dynamics constrained by isotropic dynamic facilitation. A site needs
at least one neighbour with $n_{i}=1$ to change state, with
Boltzmann probability. The concentration of excited sites is
$c=\langle n_{i}\rangle = [1+ e^{1/T}]^{-1}$, $T$ being the
temperature. Lengths are in units of the lattice spacing
which we set to one.

Self-diffusion is studied by introducing a probe molecule~\cite{jung},
which makes it possible to map `spins' back to `particles'. The
probe's position, $x(t)$, evolves with reduced time as $x(t+1)=x(t)\pm
n_{x}(t)n_{x\pm 1}(t)$, i.e., the probe is allowed to move only when
sitting on an excitation and jumps to a neighbouring excited site. Due
to coarse-graining, fast vibrations are ignored. In that sense, our
results are close to dynamic studies of inherent
structures~\cite{schroder,liao}, but this is not crucial as far as the
long-time relaxation is concerned. Numerically, we studied the
dynamics of probe molecules in the 1D FA model using Monte Carlo
simulations. At each temperature, the system size, $L$, is chosen to
be much larger than the mean distance between excitations, $\ell
(T)=1/c(T)$, in order to avoid finite size effects. We typically used
$L>10\ell (T)$ and averaged over at least $10^{4}$ independent probe
trajectories. We have also studied the behavior of a probe in the East
model~\cite{east}, a more constrained and fragile counterpart of the
1D FA model, and results are briefly discussed near the end of the
paper.

The trajectories in Fig.~\ref{bubble} show the typical time evolution
of the position of a probe molecule in the 1D FA model at a fixed, low
temperature.
Due to the kinetic constraint, excitations in dynamically
facilitated models propagate as continuous
excitation lines in space and time~\cite{jung,garrahan-chandler}. 
On the $\alpha$-relaxation timescale
motion is non-Fickian with long periods where the molecule is immobile,
seen as plateaux in Fig.~\ref{bubble}, punctuated by
shorter periods where it travels over a few sites. 
Plateaux correspond to the particle being `caged' in regions of space far from
excitations, and having to wait for an excitation line to reach its
position and allow it to move. As discussed in Ref.~\cite{jung}, the
average time it takes a probe to be reached by an excitation line and
thus move \emph{for the first time} is the mean persistence time,
which is also the $\alpha $-relaxation time, $\tau _{\alpha }$, of
these models.  Further, once the particle has moved, subsequent steps
occur when it is hit again by excitation lines, i.e., each time there
is a microscopic \emph{exchange} event. The average exchange time,
$\tau _{x}$, determines the self-diffusion constant, $D$. Since $\tau
_{x}$ grows much more slowly with decreasing temperature than $\tau
_{\alpha }$, the breakdown of the Stokes-Einstein relation
follows~\cite{jung}. This means also that the propagation of the
particle is given by a competition between these two fundamental
processes: persistence and exchange. Which one dominates at a given
temperature will depend on the time and length scales of motion.

\begin{figure}[tbp]
\begin{center}
\psfig{file=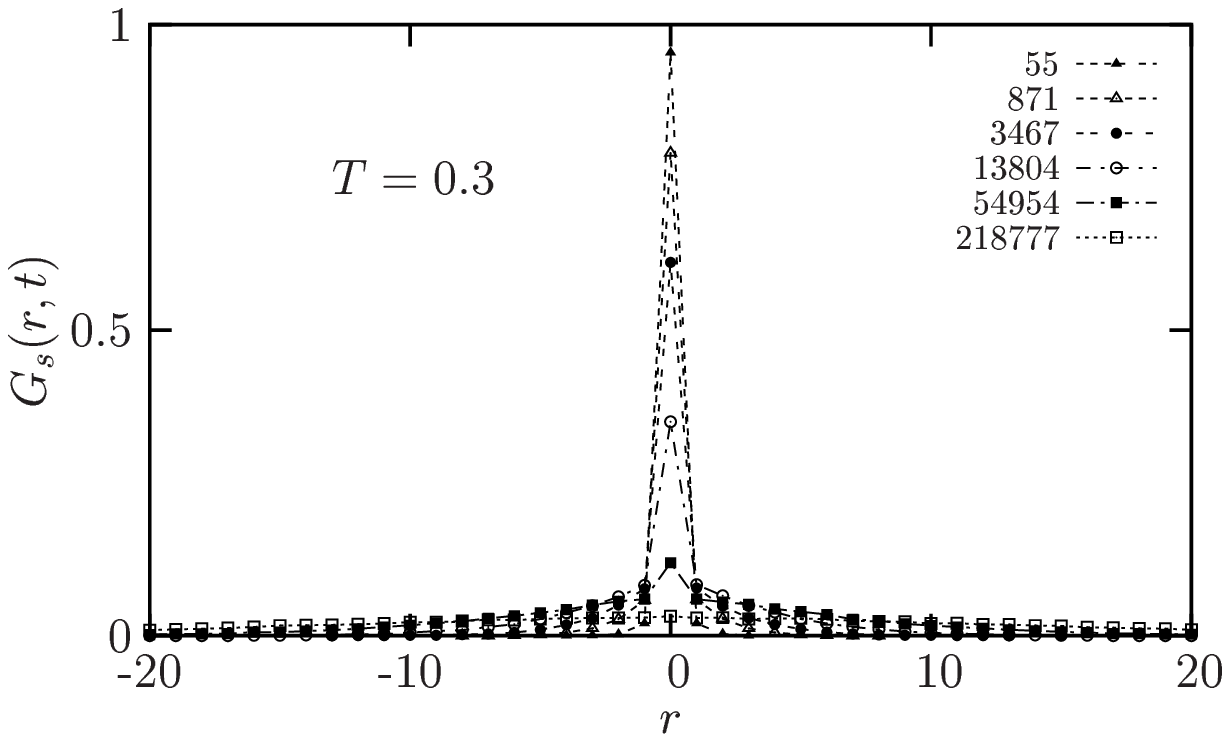,width=7.cm,height=4.5cm}
\psfig{file=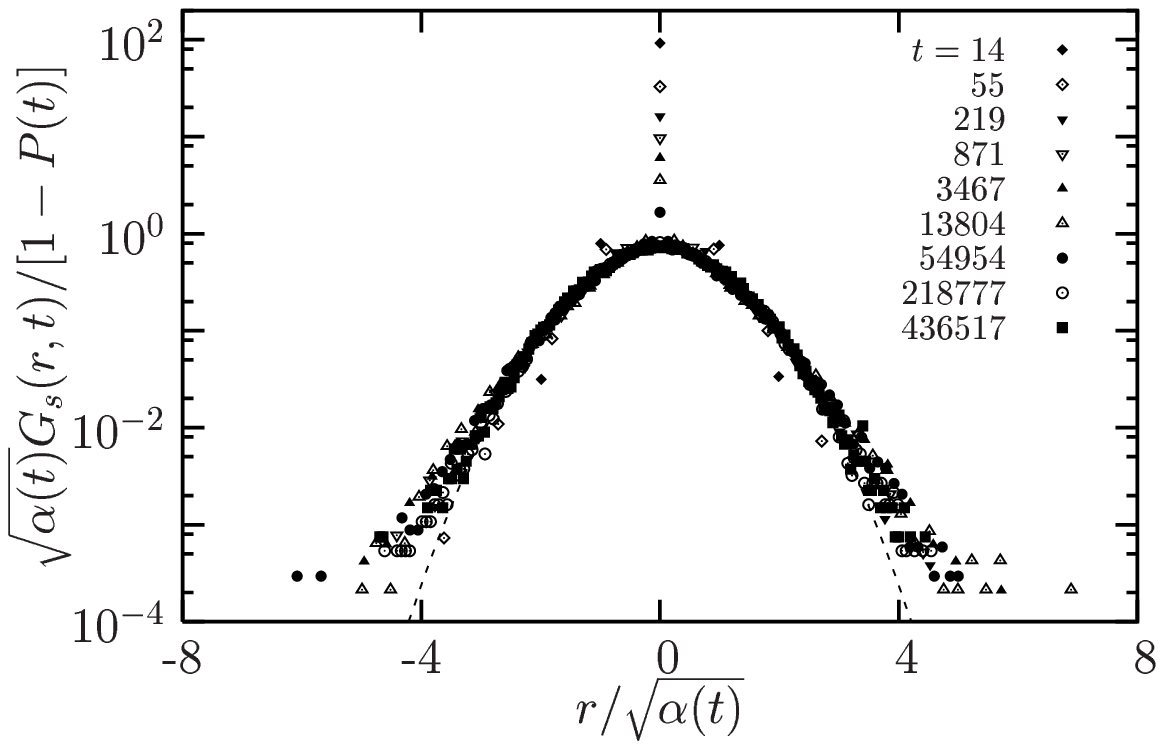,width=7.cm}
\end{center}
\caption{Left: $G_{s}(r,t)$, Eq.~(\ref{gs_eq}), at fixed temperature and
various times has a bimodal form from slow/fast molecules. Right: Diffusive
part of $G_{s}(r,t)$ rescaled after Eq.~(\ref{ansatz}) shows decreasing
deviations from the Gaussian (dashed line) with increasing time.}
\label{gs_fig}
\end{figure}

To quantify the above observations, we study the histogram of particle
displacements, i.e. the self-part of the van Hove function,
\begin{equation}
G_s(r,t) = \Big\langle \delta \big( r - [x(t) - x(0)] \big) \Big\rangle,
\label{gs_eq}
\end{equation}
where brackets indicate an average over trajectories. The above
discussion suggests that at times of the order of $\tau_\alpha$, some
particles are still trapped in their initial position, while others
have already escaped and rapidly diffused over a certain distance. The
distribution (\ref{gs_eq}) should reflect this bimodal character with
a contribution in $r=0$ from the `slow', immobile particles, and
contributions at non-zero displacements from diffusing, `fast'
particles. Numerical results in Fig.~\ref{gs_fig} (left panel) fully
confirm this expectation. Coexistence of fast and slow subpopulations
of particles is a major indicator of dynamic
heterogeneity~\cite{DHreviews1,DHreviews2,DHreviews3,DHreviews4,van,kob1,weeks}.

The behavior of the van Hove function (\ref{gs_eq}) can be anticipated
from a simple analysis. Each time the probe particle is allowed to
move it makes a random walk step. Therefore, it performs a random
walk, but the time lag between steps fluctuates. The time lags are
determined by the dynamics of the mobility field of the host
liquid. For ease of analysis, we neglect the back reaction of the
probe on this field. In this approximation, the probe process is a
random walk at random times, or continuous time random
walk~\cite{Montroll-Weiss}. The probability for the particle to be at
position $r$ at time $t$ is then $G_{s}(r,t)=\sum_{m=0}^{\infty }\pi
_{m}(t)\phi ^{(m)}(r) $, where $\pi _{m}(t)$ is the probability that
the probe jumped $m$ times up to time $t$, and $\phi ^{(m)}(r)$ is the
probability that a random walker is at site $r$ after $m$ steps,
having started from the origin.

The first step is a persistence event. If $p(t)$ is the distribution
of persistence times \cite{pre}, then the probability $\pi_0(t)$ of
not having an event up to time $t$ is given by $\pi_0(t) = P(t)$, where
$P(t)=\int_t^\infty dt^{\prime}p(t^{\prime})$ is the persistence
function \cite{garrahan-chandler,pre}. If we further assume that
successive exchange events are uncorrelated, which is a good
approximation in the FA model \cite {jung}, then $\pi_m(t)$ can be
written as a multiple convolution of the exchange time distribution
$\psi(t)$ \cite{grimmet}. The Laplace transform of $\pi_m(t)$ for $m >
0$ then reads: $\hat{\pi}_m(s) = \hat{p}(s) \hat{\psi}^{m-1}(s)
[1-\hat{\psi}(s)]/s$, where hats indicate Laplace transforms. If we
also Fourier transform in space we obtain $\hat{F}_s(k,s)$, which is
the Laplace transform of the self-intermediate scattering function,
$F_s(k,t)$, and reads
\begin{equation}
\hat{F}_s(k,s) = \hat{P}(s) + \cos{(k)} \frac{\hat{p}(s)}{s} \frac{ 1
- \hat{\psi}(s)}{ 1 - \cos{(k)} \hat{\psi}(s)} \; .  \label{MWeq}
\end{equation}
This is the Montroll-Weiss equation for the propagator of the probe
particle~\cite{Montroll-Weiss}. From Eq.~(\ref{MWeq}) we immediately
obtain a $k$ and $T$ dependent timescale: $\tau(k,T) = \lim_{s \to 0}
\hat{F}_s(k,s)$. Since the distributions $p(t)$ and $\psi(t)$ are both
narrow (all moments are finite~\cite{garrahan-chandler}), 
their Laplace transforms read, for
small $s$, $\hat{p}(s) \approx 1 - \tau_\alpha s$ and $\hat{\psi}(s)
\approx 1 - \tau_x s$, which leads to:
\begin{equation}
\tau(k,T) = \tau_\alpha(T) + \frac{\cos{(k)}}{1-\cos{(k)}} \tau_x(T) \; .
\label{tauk}
\end{equation}
For small $k$ this gives $\tau \approx \tau_\alpha + (k^2 D)^{-1}$, since $D
= 1/(2 \tau_x)$ \cite{jung}.

The inverse Laplace transform of the factor $\cos
{(k)}[1-\hat{\psi}][s-s\cos {(k)}\hat{\psi}]^{-1}$ in Eq.\
(\ref{MWeq}) defines a propagator for diffusing molecules,
$F_{\mathrm{diff}}(k,t)$. Due to the narrowness of $\psi $, it
becomes that of normal diffusion in the large $t$, small $k$ limit,
$F_{\mathrm{diff}}(k,t)\approx \exp {(-Dk^{2}t)}$. Inverse Laplace
transforming (\ref{MWeq}) we obtain
$F_{s}(k,t)=P(t)+\int_{0}^{t}dt^{\prime }p(t^{\prime
})F_{\mathrm{diff}}(k,t-t^{\prime })$. In the regime of strong
decoupling this can be further approximated by:
\begin{equation}
F_{s}(k,t)\approx P(t)+\left[ 1-P(t)\right] F_{\mathrm{diff}}(k,t)\;.
\label{fsk}
\end{equation}
The corresponding van Hove function then reads,
\begin{equation}
G_{s}(r,t)\approx P(t)\delta (r)+\frac{1-P(t)}{\sqrt{2\pi \alpha
(t)}}\exp {\left( -\frac{r^{2}}{2\alpha (t)}\right) }\;,
\label{ansatz}
\end{equation}
where we have approximated $G_{\mathrm{diff}}(r,t)$ by a Gaussian for all
times. We find that Eq.\ (\ref{ansatz}) works well for all times and
temperatures, see Fig.\ \ref{gs_fig} (right). Deviations from
the Gaussian tend to disappear at long times, as expected. The mean
square displacement reads $\langle r^{2}\rangle =\int_{-\infty }^{\infty
}dr\,r^{2}G_{s}(r,t)\simeq \big[ 1-P(t)\big]
\alpha (t)$, where $\alpha (t)$ is the mean square displacement obtained by
restricting the average to the fast particles. For large times, $\alpha
(t\to \infty )\approx \langle r^{2}\rangle \approx 2Dt$, while $\alpha (t\to
0)=1$ by construction. Physically, Eq.~(\ref{ansatz}) shows that it is a
good approximation to think of molecules as existing in one of two
subpopulations of immobile and mobile particles.

Our primary results in this paper all follow from Eq.\ (\ref{ansatz}).
It is valid only for lengths and times larger than our lattice spacing
and unit of time, respectively.  Thus the delta-function in Eq.\
(\ref{ansatz}) contributes to an experimentally observed $G_{s}(r,t)$
only after coarse-graining and would be replaced at smaller 
length scales by a smoother distribution, typically Gaussian. 
A two-Gaussian
fit to the van Hove function is reported in confocal microscopy experiments
performed with colloidal suspensions~\cite{van}. If one is predisposed to
think of the distribution in terms of a single Gaussian, it will appear as
if the distribution has `fat tails.' Such tails have been described as
indicators of dynamic heterogeneity~\cite{kob1,DHreviews2,weeks}.

\begin{figure}[tbp]
\begin{center}
\psfig{file=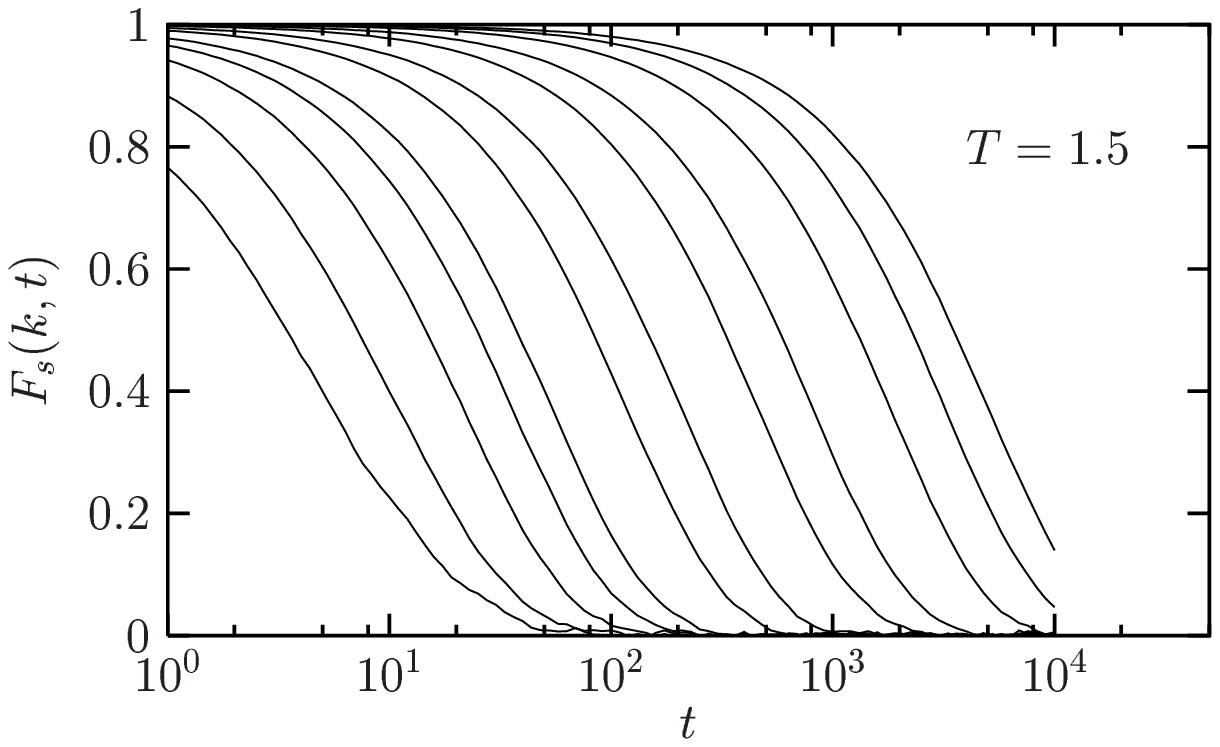,width=6.8cm} \psfig{file=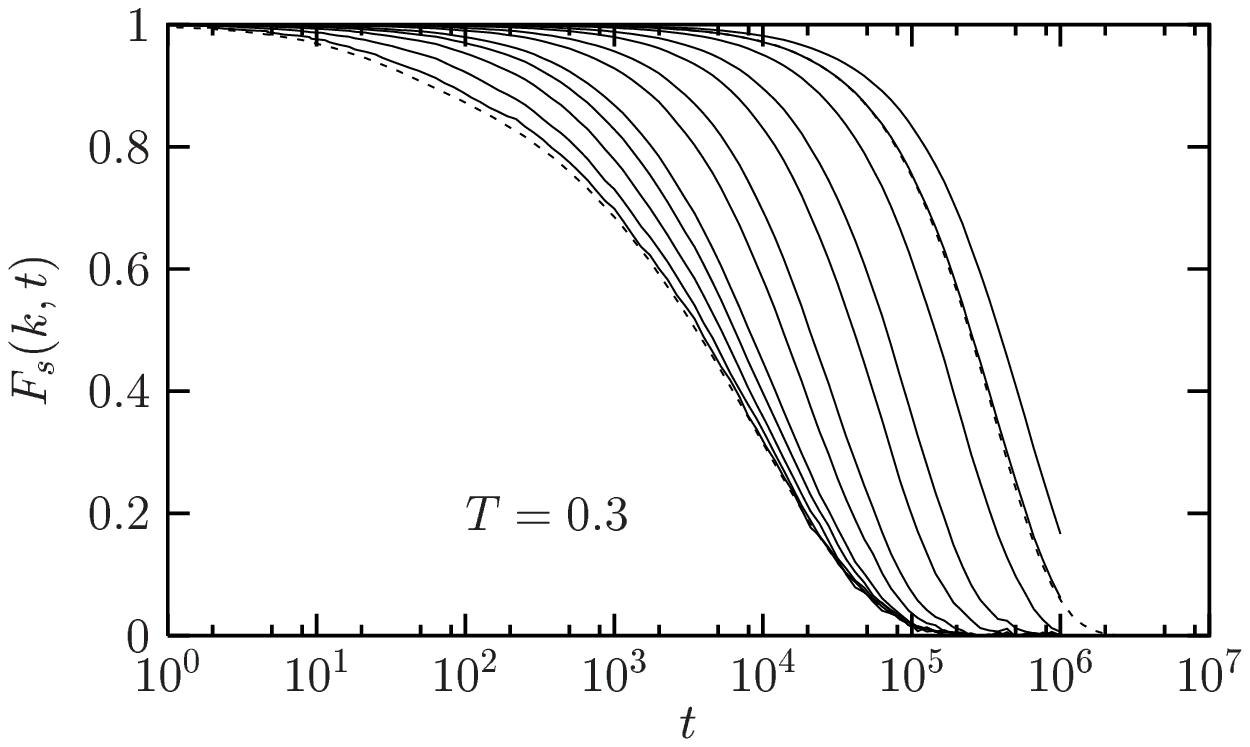,width=6.8cm} \\[0pt]
\psfig{file=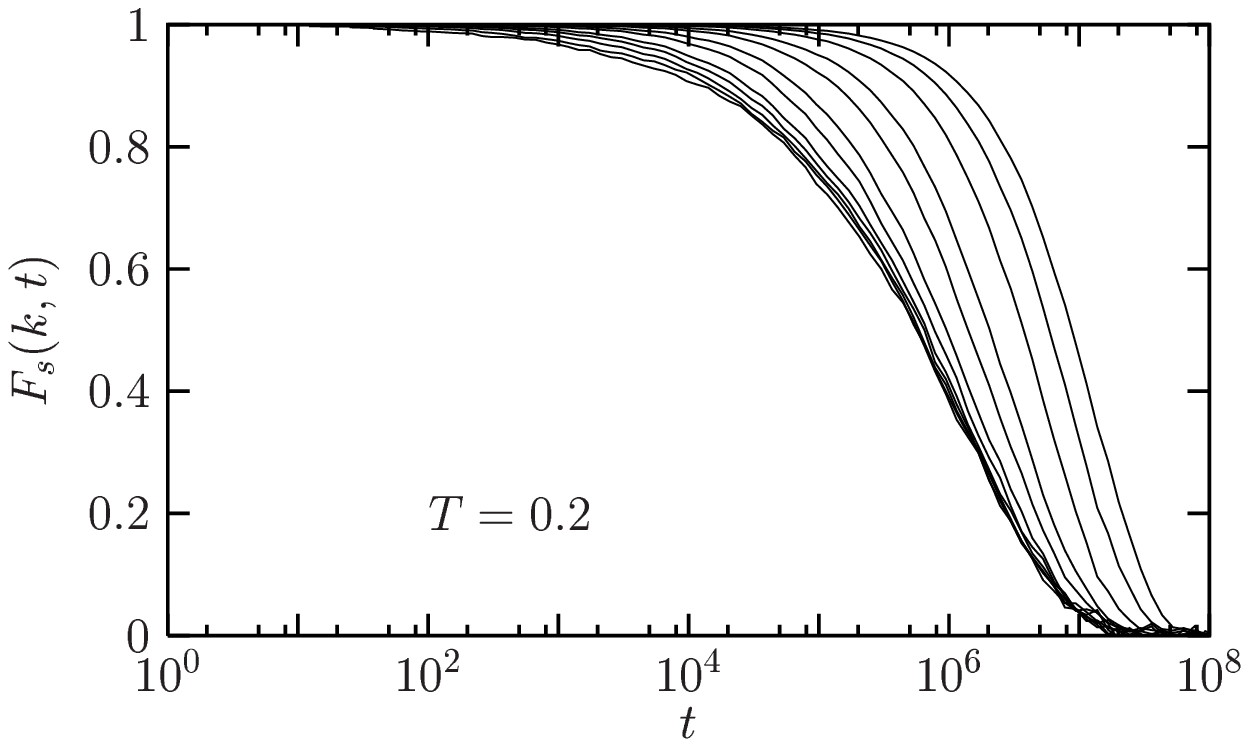,width=6.8cm} \psfig{file=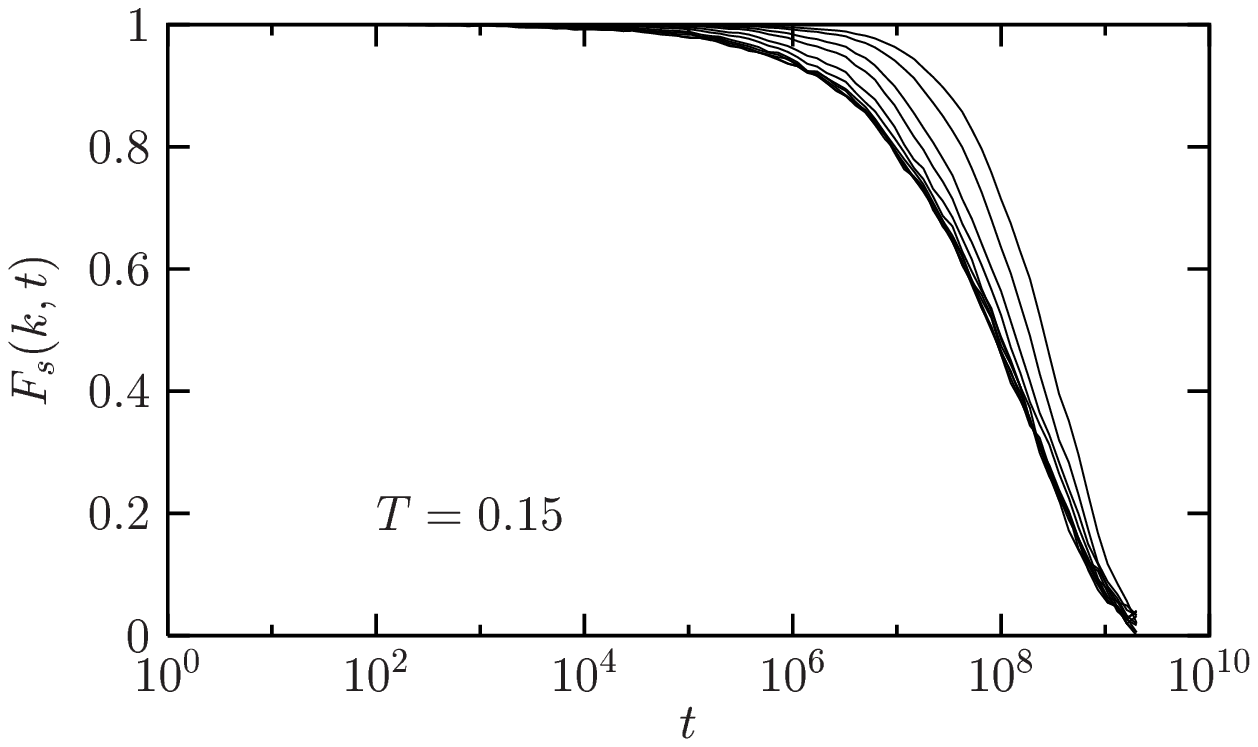,width=6.8cm}
\end{center}
\caption{Self-intermediate scattering function in the 1D FA model, for
different $T$ and the same range of $k$: $k=2 \pi /n$, with $n=2$, 4,
6, 8, 10, 15, 20, 30, 40, 60, 80, and 100 (from left to right). Dashed
lines at $T=0.3$ represent the persistence $P(t)$ and the diffusive
behaviour $\exp[-(2 \pi / 80)^2 Dt]$.}
\label{fs}
\end{figure}

The approximation (\ref{ansatz}) also provides a simple explanation
for the time dependence of the non-Gaussian parameter,
$\alpha_{2}(t)=\frac{1}{3}\langle r^{4}\rangle /\langle r^{2}\rangle
$. By construction, $\alpha _{2}(t)=1$ for a Gaussian
$G_{s}(r,t)$. Instead, $\alpha _{2}(t)$ reflects the transition from a
Gaussian (fast vibrations at short times) to another Gaussian (Fickian
diffusion at long times), the distribution being non-Gaussian at
intermediate times when both terms in (\ref{ansatz}) compete. From
Eq.\ (\ref{ansatz}) we obtain $\alpha _{2}(t)=\big[ 1-P(t)\big]^{-1}$,
which is monotonically decreasing with time.  This expression is valid
beyond a coarse graining time, and therefore does not describe the
very short-time Gaussian behaviour that would make $\alpha _{2}(t)$
non-monotonic.  
A similar monotonic behaviour is found in molecular 
dynamics studies of inherent structures dynamics \cite{liao}.
This expression does, however, explain 
why $\alpha_{2}(t)$ can reach values much larger than unity.

Figure~\ref{fs} presents our numerical results at various temperatures
and wavevectors for the self-intermediate scattering function
$F_s(k,t)$. At high temperature, $T=1.5$, relaxation is homogeneous
and $F_s(k,t)$ decays exponentially at all wavevectors. The effect of
dynamic heterogeneity becomes fully visible when $T$ decreases. First,
time decays are non-exponential at large wavevector. Second,
$F_s(k,t)$ becomes exponential when $k$ decreases at constant $T$, as
suggested by our observations of Fig.~\ref{bubble}. Finally, curves
for different wavevectors that are distinct at high $T$ superpose when
$T$ is decreased, and $F_s(k,t)$ becomes $k$-independent.

These observations are rationalized by Eq.\ (\ref{fsk}). For large $k$, the
last term in (\ref{fsk}) is exponentially suppressed, and $F_{s}(k,t)\approx
P(t)$, which is $\exp (-\sqrt{t/\tau _{\alpha }})$ in the 1D FA 
model~\cite{note}. This behaviour justifies our identification of 
$\tau_\alpha$ with the persistence time~\cite{jung,pre}.   
At small $k$ and large times, Eq.~(\ref{fsk}) becomes
$F_{s}(k,t)\approx \exp (-k^{2}Dt)$, as confirmed by Fig.~\ref{fs}. At
intermediate $k$, mixed behaviour between diffusion and persistence is
predicted, and observed. These observations naturally imply that stretching
is also a $k$ and $T$ dependent property, at odds with predictions stemming
from mode-coupling theory~\cite{mct}, which indeed poorly describe
self-dynamics at intermediate wavevectors~\cite{mct}.

\begin{figure}[tbp]
\begin{center}
\psfig{file=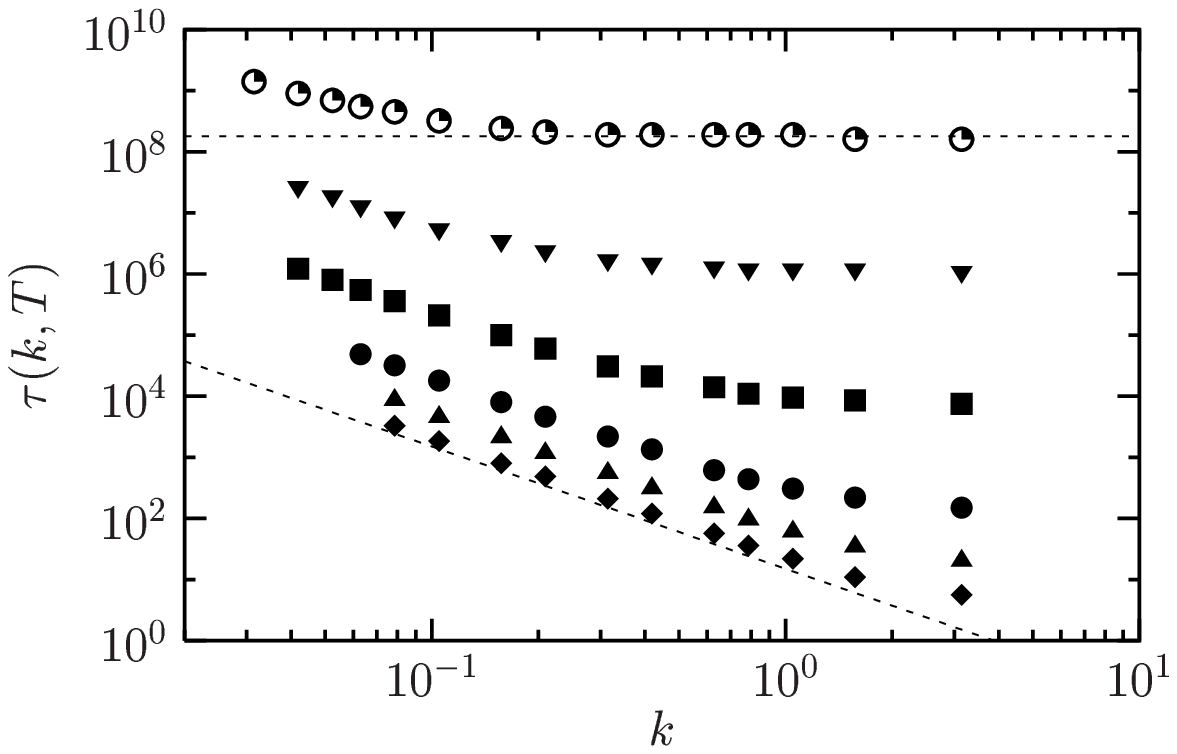,width=7.cm} \psfig{file=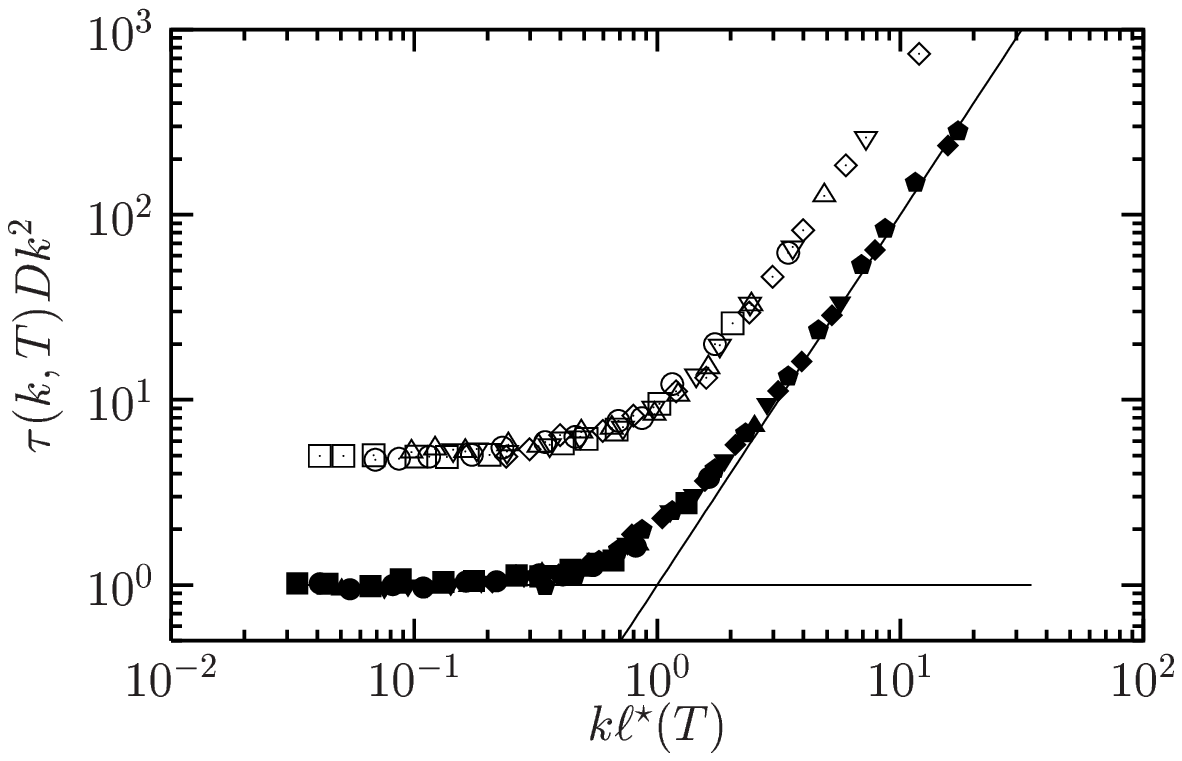,width=7.cm}
\end{center}
\caption{Left: Relaxation times in the 1D FA model, as a function of
wavevector for different temperatures, $T=1.5$, 0.8, 0.6, 0.3, 0.2,
and 0.15 (from bottom to top). The straight lines correspond to $\tau
\sim k^{-2}$ and $\tau \sim const$, the limiting behaviours predicted
by Eq.\ (\ref{tauk}). Right: timescales rescaled by the diffusive
limit $k^{-2} D^{-1}$ are collapsed using the scaled variable $k
\ell^\star$ for both 1D FA (filled symbols) and East (open symbols)
models. Data for the East model cover about 10 decades in timescales,
with $T \in [0.31, 0.8]$ and the scaling curve has been shifted by a
factor 5, for clarity.}
\label{time}
\end{figure}

The crossover from persistent to Fickian diffusive dynamics shows up
in the behaviour of the relaxation times $\tau (k,T)$, as given by
Eq. (\ref{tauk}). Numerically, we extract the timescales from the
scattering functions in the usual way, $F_{s}(k,\tau
)=e^{-1}$. Results are presented in Fig.~\ref {time}. Fickian
behaviour, $\tau \approx k^{-2}D^{-1}$, is obeyed at all $k$ at high
$T$. When $T$ decreases, diffusive behaviour is restricted to smaller
and smaller $k$, and is progressively replaced, at larger $k$, by the
$k$-independent form $\tau \approx \tau _{\alpha }$. This leads us to
define a temperature dependent characteristic length scale, $\ell
^{\star }(T)$, which determines the onset of Fickian diffusion. From
Eq.\ (\ref{tauk}) we get
\begin{equation}
\ell ^{\star }\sim \sqrt{D\tau _{\alpha }}\;.  \label{main}
\end{equation}
It is therefore possible to collapse all timescales
on a master curve rescaling times by $k^{-2}D^{-1}$ and space by
$\ell ^{\star }$. This is shown in the right panel of
Fig.~\ref{time}, where $\ell ^{\star }$ is estimated from
Eq.~(\ref{main}). In Fig.\ \ref {time}, we also show a similar
collapse of time scales that we have found when the East
model~\cite{east} is used in place of the 1D FA model. The East model
is the fragile counterpart of the 1D FA model. We see that the scaling
behaviour predicted by Eqs.\ (\ref{tauk}, \ref{main}) works well also
in this case, despite the fact that in the East model successive
exchange events are not uncorrelated \cite{jung}, which was one of the
assumptions in the derivation of Eqs.\ (\ref{MWeq})-(\ref{ansatz}). A
behaviour similar to that of Figs.~\ref{fs} and \ref{time} was
recently reported in molecular dynamics simulations of a binary
Lennard-Jones mixture~\cite{ludo}, and in experiments on supercooled
TNB~\cite{mark}.

For both the FA and East models
the diffusion constant obeys a fractional Stokes-Einstein law, $D\sim
\tau _{\alpha }^{-\xi }$, with $\xi \leq 1$ \cite{jung}. In the case
of the FA model, $\xi _{\mathrm{FA}}=2/\Delta \approx 2/3$, $2/2.3$,
$2/2.1$ for dimensions $d=1$, 2, 3, respectively, $\Delta $ being the
time exponent $\tau _{\alpha }\sim c^{-\Delta }$
\cite{steve1,steve2}. The Stokes-Einstein law, $\xi =1$, is recovered
for $d\geq 4$, the upper critical dimension of the FA model
\cite{steve1,steve2}. For the East model numerical results indicate
$\xi _{\mathrm{East}}\approx 0.7-0.8,$ independent of $d$ up to the
highest dimensionality studied, $d\leq 6$. Equation (\ref{main}) can
be rewritten,
\begin{equation}
\ell ^{\star }\sim \tau _{\alpha }^{(1-\xi )/2}\;.  \label{lst}
\end{equation}
Therefore, $\ell ^{\star }$ will diverge when $T\to 0$ (or
$c\rightarrow 0)$ if there is Stokes-Einstein breakdown, i.e., when
$\xi <1$.  
The Fickian crossover length, $\ell ^{\star},$ measures something
related to but distinct from the largest dynamic heterogeneity length,
$\ell (T)$, which can be measured for example through multi-point
dynamic structure factors~\cite{DHreviews3}. For both strong (FA) and
fragile (East) systems, this length goes as $\ell \sim c^{-\nu }$ with
a spatial exponent $\nu $~\cite{garrahan-chandler,steve1}, and
therefore always diverges as $T\to 0$ in an Arrhenius manner. In
general, we have that $\ell \neq \ell ^{\star }$, and in particular
for fragile systems, Eq.\ (\ref{lst}) shows that $\ell ^{\star }$ will
grow faster than $\ell $, in a super-Arrhenius way. 
The reverse is true in strong systems where $\ell$ 
should grow faster than $\ell^\star$.
In Ref.~\cite {ludo},
the typical length scale $\ell $ of dynamic heterogeneity was used to
rescale wavevectors in the analog of Fig.~\ref{time}. Present results
show instead that $\ell $ and $\ell ^{\star }$ are different
quantities, although they might be hard to distinguish on a restricted
temperature window.
Finally, $\ell^\star$
was identified in Ref.~\cite{colby} by dimensional
analysis which confuses $\ell$ and $\ell^\star$. This particular confusion is
avoided in Ref.~\cite{Schweizer} where the crossover length 
is also obtained, by assuming a memory function is
analytic for small wave vectors and its Taylor series can be
truncated at second order even for wave-vectors that are not small.

The physical basis for the Fickian crossover is that persistence time
dominates over exchange time. The former is the time for the first
dynamical step, while the latter is the typical time scale of
subsequent motion.  This mechanism naturally leads to the breakdown of
the Stokes-Einstein law~\cite{jung}, non-Gaussian van Hove functions
with large tails, and sub-populations of fast and slow particles.  The
present picture is different from the idea that relaxation and
diffusion result from averaging a time and its inverse, respectively,
over a single distribution of relaxation times~\cite
{Schweizer,gilles}. Our explanation that physically distinct processes
(persistence and exchange) compete is consistent with the experimental
observation~\cite{mark} of a strong decoupling in a material with
self-similar distributions of relaxation times.

\acknowledgments We are grateful to G.~Biroli,
Y.~Jung, and S.~Whitelam for discussions, and 
to M.D.~Ediger and K.~Schweizer for
discussions about their published and unpublished results.  
This work was supported by
EPSRC Grants No.\ GR/R83712/01 and GR/S54074/01, E.U.\
Marie Curie Grant No.\ HPMF-CT-2002-01927, University of Nottingham
Grant No.\ FEF 3024, the US National Science
Foundation, and Oxford Supercomputing Center. Much of this work was
carried out in February, 2004, when DC was a Schlumberger Visiting
Professor at Oxford.


\begin{thebibliography}{99}
\bibitem{note} The functional form of $P(t)$ is dimensional dependent.
See, for example, \cite{steve2}.
\end{thebibliography}
\end{document}